\documentclass[12pt]{article}

\usepackage{amssymb}
\usepackage{amsmath}
\usepackage{graphicx}
\usepackage{bm}

\newcommand{\INTM}{\int\frac{d^3\bm p}{(2\pi)^3}}

\def\beq{\begin{equation}}
\def\eeq#1{\label{#1}\end{equation}}
\def\eeqn{\end{equation}}

\def\beqa{\begin{eqnarray}}
\def\eeqa#1{\label{#1}\end{eqnarray}}
\def\eeqan{\end{eqnarray}}

\let\bar=\overbar

\def\Dslash{\not{\hbox{\kern-4pt $D$}}}
\def\dslash{\not{\hbox{\kern-2pt $\del$}}}

\def\msb{{\bar{\ssstyle M \kern -1pt S}}}

\def\Title#1{\begin{center} {\Large {\bf #1} } \end{center}}

\begin{document}

\Title{Quark Matter with a Chiral Chemical Potential}

\bigskip\bigskip

%+\addtocontents{toc}{{\it D. Reggiano}}
%+\label{ReggianoStart}

\begin{raggedright}  

{\it Marco Ruggieri\index{Ruggieri, M.}\\
Department of Physics and Astronomy, University of Catania, Via S. Sofia 64, I-95125 Catania}
\bigskip\bigskip
\end{raggedright}

\section{Introduction}
In this talk, we report the results discussed in~\cite{Ruggieri:2011xc},
related to the phase structure of hot quark matter in presence of a background
of chiral charge density, $n_5$. The latter is introduced in the simplest way possible,
namely by virtue of a chemical potential, $\mu_5$, conjugated
to $n_5$. In more detail, after a brief introduction and list
of motivations of this kind of study, we discuss the interplay between
chiral symmetry restoration and deconfinement at finite $\mu_5$,
as well as the critical endpoint in the phase diagram and its possible
relationship with the critical endpoint of the phase diagram od
Quantum Chromodynamics (QCD).  In the talk, due to time limitation,
we can emphasize few results related to the latter topic. Therefore we need
to leave apart several applications of the ideas, as well as of the formalism,
developed here to the physics of heavy ion collisions, with particular reference
to the Chiral Magnetic Effect~\cite{Kharzeev:2007jp,Fukushima:2008xe,Buividovich:2009wi}. 
The latter might be relevant for the phenomenology of
heavy ion collisions, because a copious production of gluon configurations,
the QCD sphalerons, with a finite winding number is expected in the quark-gluon-plasma
phase of QCD, see~\cite{Moore:2010jd} and references therein. Because
of the chiral Ward identity, the interaction of the sphalerons 
with the quarks causes a chirality change of the latters. As a consequence,
a copious production of local domains in which chirality is imbalanced,
is expected in the quark-gluon-plasma.

The critical endpoint, CP, of QCD~\cite{Asakawa:1989bq} 
is the cornerstone of the phase diagram of strongly interacting matter. 
At CP, a crossover line and a first order line are supposed to intercept. 
It is thus
not surprising that an intense experimental activity is nowadays
dedicated to the detection of such a point, which involves the
large facilities at RHIC and LHC; moreover, further experiments
are expected after the development of FAIR at GSI. Several
theoretical signatures of CP have been
suggested~\cite{Sig-CEP,Stephanov:1999zu}. Despite the importance
of CP, a firm theoretical evidence of its existence is still
missing. In fact, the sign problem makes the Lattice Monte Carlo
simulations difficult, if not impossible, in the large
baryon-chemical potential ($\mu$) region for 
$N_c=3$~\cite{LQCD-CEP}, see~\cite{deForcrand:2010ys} for a recent
review. Therefore, it has not yet been possible to prove
unambiguously the existence and the location of CP starting from
first principles simulations of grand-canonical ensembles.  Moreover,
the predictions of effective models are spread in the $T-\mu$
plane, see for example~\cite{Stephanov:2007fk,Ohnishi:2011jv}.

Interesting overcomings of the sign problem for the quest of CP
are: analytic continuation of data obtained at imaginary
chemical potential, $\mu_I$~\cite{Alford:1998sd,de
Forcrand:2002ci,D'Elia:2002gd,deForcrand:2008vr}; 
simulations at finite isospin chemical potential, see for
example~\cite{Kogut:2002zg,deForcrand:2007uz,Cea:2009ba}; simulations in canonical, rather
then grand-canonical, ensembles~\cite{Li:2011ee}; 
strong coupling expansion of Lattice
QCD~\cite{SC-LQCD,deForcrand:2009dh}.
On the
purely theoretical side, it has been suggested~\cite{Hanada:2011ju} 
that the use of orbifold equivalence
in the large $N_c$ approximation of QCD can lead to relations
between the coordinates of CP at finite chemical potential, with
those at finite isospin chemical potential.

In this talk, we present the idea suggested in~\cite{Ruggieri:2011xc} 
about a new theoretical way to detect the
CP, by means of Lattice simulations with $N_c = 3$. In order to
accomplish this important program, we suggest to simulate QCD with
a chiral chemical potential, $\mu_5$, conjugated to the chiral
charge density, $n_5 = n_R - n_L$,
see~\cite{Fukushima:2008xe,Fukushima:2010fe,Chernodub:2011fr,McLerran:1990de,Bayona:2011ab}
for previous studies. Our idea, supported by concrete calculations
within acmicroscopic effective model, is that CP can be
continued to a critical endpoint at $\mu_5 \neq 0$ and $\mu = 0$,
that we denote by CP$_5$, the latter being accessible to $N_c = 3$
Lattice QCD simulations of grand-canonical
ensembles~\cite{Fukushima:2008xe,Yamamoto:2011gk}. Therefore, the detection of the
former endpoint via Lattice simulations, can be considered as a
signal of the existence of the latter. 

The model used in the calculation, namely the
Nambu-Jona-Lasinio model with the Polyakov
loop~\cite{Fukushima:2003fw} (PNJL model in the following) with
tree level coupling among chiral condensate and Polyakov
loop~\cite{Sakai:2010rp}, gives numerical relations among the
coordinates of CP$_5$ and those of CP. In particular, the critical
temperature turns out to be almost unaffected by the process of
continuation; the critical value of the chemical potential,
$\mu_c$, on the other hand turns out to be almost half of the
critical chiral chemical potential, $\mu_{5c}$.

Before discussing our results, it is important to spend some word
more about the chiral chemical potential. In particular, we are
aware that world at finite $\mu_5$ should be considered as a fictional
one. As a matter of fact, $\mu_5$ cannot be considered as a
true chemical potential because, in the confinement phase, the
chiral condensate $\langle\bar qq\rangle$ mixes left- and
right-handed components of the quark field, leading to
non-conservation of $n_5$. Moreover, the quantum chiral anomaly
leads to fluctuations of the topological charge, which in turn 
causes the changes of the chiral density because of the Ward identity. 
Therefore, the point of view that we adopt is to consider
$\mu_5$ as a mere mathematical artifice. However, the world at finite
$\mu_5$ with $N_c = 3$ can be simulated on the Lattice. Therefore, it is worth to study it by
grand-canonical ensemble simulations: it might furnish an evidence
of the existence of the critical endpoint in the real world.

\section{The Model with the Polyakov loop}
Because of its non-perturbative nature, we cannot make first
principles calculations within QCD in the regimes to which we are
interested in, namely moderate $T$, $\mu$ and $\mu_5$. Hence we
need to rely on some effective model, which is built in order to
respect (at least some of) the symmetries of the QCD action. To
this end, we make use of the Nambu-Jona-Lasinio
model~\cite{Nambu:1961tp} (see~\cite{revNJL} for reviews) improved
with the Polyakov loop~\cite{Fukushima:2003fw}, dubbed PNJL model,
which has been used many times in recent years to describe
successfully the thermodynamics of QCD with two and two-plus-one
flavors,
see~\cite{Sakai:2010rp,Roessner:2006xn,Sasaki:2006ww,Abuki:2008nm,
Kashiwa:2007hw,Herbst:2010rf,Kahara:2008yg,Skokov:2010uh,Andersen:2011pr}
and references therein. The model is interesting because it allows
for a self-consistent description of spontaneous chiral symmetry
breaking; even more, it allows for a simultaneous computation of
quantities sensible to confinement and chiral symmetry breaking.
We restrict here to a brief summary of the main equations; we refer
to~\cite{Ruggieri:2011xc} for a more detailed discussion.

In the PNJL model, quark propagation in the medium is described by
the following lagrangian density:
\begin{equation}
{\cal L} =\bar q\left(i\gamma^\mu D_\mu - m + \mu_5 \gamma^0\gamma^5 +
\mu\gamma^0 \right)q + G\left[\left(\bar qq\right)^2 + \left(i\bar
q\gamma_5\bm\tau q\right)^2\right]~;\label{eq:1ooo}
\end{equation}
In the above equation, $q$ corresponds to a quark field in the
fundamental representation of color group $SU(3)$ and flavor group
$SU(2)$. We have a introduced chemical potential for the quark number density,
$\mu$, and a pseudo-chemical potential conjugated to chirality
imbalance, $\mu_5$.  The chiral charge density, $n_5 = n_R -n_L$, 
represents the difference in densities of the right- and left-handed quarks. 
The imbalance of chiral density
can be created by instanton/sphaleron transition in QCD,
see~\cite{Fukushima:2008xe} and references therein.
At finite $\mu_5$, a chirality imbalance is created, namely $n_5 \neq 0$. For
example, in the massless limit and at zero baryon chemical
potential one has~\cite{Fukushima:2008xe}
\begin{equation}
n_5 = \frac{\mu_5^3}{3\pi^2} + \frac{\mu_5 T^2}{3}~.
\end{equation}
If quark mass (bare or constituent) is taken into account, the
relation $n_5(\mu_5)$ cannot be found analytically in the general
case, and a numerical investigation is needed, see for
example~\cite{Fukushima:2010fe}.

In our computation we follow the idea implemented
in~\cite{Sakai:2010rp}, which brings to a Polyakov-loop-dependent
coupling constant:
\begin{equation}
G = g\left[1 - \alpha_1 L L^\dagger -\alpha_2(L^3 +
(L^\dagger)^3)\right]~,\label{eq:Run}
\end{equation}
The ansatz in the above equation was inspired
by~\cite{Kondo:2010ts,Frasca:2008zp} in which it was shown
explicitly that the NJL vertex can be derived in the infrared
limit of QCD, it has a non-local structure, and it acquires a
non-trivial dependence on the phase of the Polyakov loop. We refer
to~\cite{Sakai:2010rp} for a more detailed discussion. This idea
has been analyzed recently in~\cite{Braun:2011fw}, where the
effect of the confinement order parameter on the four-fermion
interactions and their renormalization-group fixed-point structure
has been investigated. The numerical values of $\alpha_1$ and
$\alpha_2$ have been fixed in~\cite{Sakai:2010rp} by a best fit of
the available Lattice data at zero and imaginary chemical
potential of Ref.~\cite{D'Elia:2009qz,Bonati:2010gi}. In
particular, the fitted data are the critical temperature at zero
chemical potential, and the dependence of the Roberge-Weiss
endpoint on the bare quark mass. The best fit procedure leads to
$\alpha_1 = \alpha_2 \equiv \alpha = 0.2 \pm 0.05$.

In the one-loop approximation, the effective potential of this
model is given by
\begin{eqnarray}
V &=& {\cal U}(L,L^\dagger,T) + \sigma^2 G  -N_c N_f\sum_{s=\pm 1}\INTM \omega_s \nonumber \\
&&-\frac{N_f}{\beta}\sum_{s=\pm 1}\INTM\log\left(F_+ F_-\right)
\label{eq:OB}
\end{eqnarray}
where
\begin{equation}
\omega_s = \sqrt{(|\bm p| s -\mu_5)^2 + m_q^2}~, \label{eq:iii}
\end{equation}
corresponds to the pole of the quark propagator, and
\begin{eqnarray}
F_- &=& 1+3L e^{-\beta(\omega_s - \mu)} +3L^\dagger
e^{-2\beta(\omega_s - \mu)} + e^{-3\beta(\omega_s -
\mu)}~, \\
F_+ &=& 1+3L^\dagger e^{-\beta(\omega_s + \mu)} +3L
e^{-2\beta(\omega_s + \mu)} + e^{-3\beta(\omega_s +
\mu)}~,
\end{eqnarray}
denote the statistical confining thermal contributions to the
effective potential; $\omega_s$ is given by
Equation~\eqref{eq:iii}, with $m_q = m -2 G \sigma$. Once again the
vacuum fluctuation term is regularized by means of a ultraviolet
cutoff, that we denote by $M$. The relation between the chiral
condensate and $\sigma$ in the PNJL model is $\sigma = \langle\bar qq\rangle$.

We notice that the PNJL model considered here, which is dubbed Extended-PNJL
in~\cite{Sakai:2010rp}, has been tuned in order to reproduce
quantitatively the Lattice QCD thermodynamics at zero and
imaginary quark chemical potential. Hence, it represents a
faithful description of QCD, in terms of collective degrees of
freedom related to chiral symmetry breaking and deconfinement.

The potential term $\mathcal{U}$ in Eq.~\eqref{eq:OB} is built by
hand in order to reproduce the pure gluonic lattice data with $N_c
= 3$~\cite{Roessner:2006xn}. We adopt the following logarithmic
form,
\begin{equation}
 \mathcal{U}[L,\bar L,T] = T^4\biggl\{-\frac{a(T)}{2}
  \bar L L 
 + b(T)\ln\bigl[ 1-6\bar LL + 4(\bar L^3 + L^3)
  -3(\bar LL)^2 \bigr] \biggr\}~.
\label{eq:Poly}
\end{equation}
We refer to~\cite{Roessner:2006xn,Ruggieri:2011xc} for the numerical values 
of the parameters used in this study.

\section{Critical endpoint at zero chemical potential}

\begin{figure}[t!]
\begin{center}
\includegraphics[width=8.5cm]{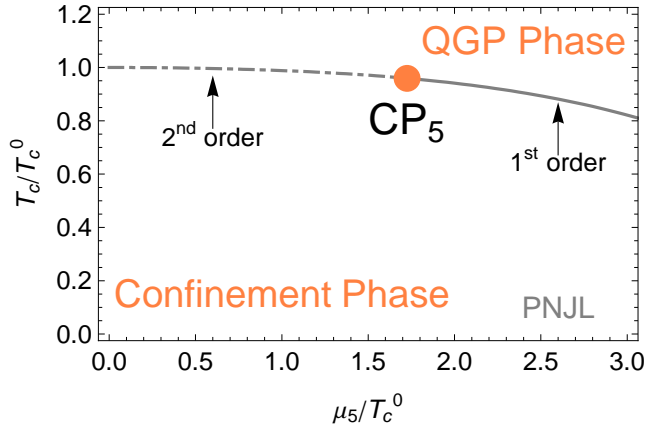}
%\end{center}
\caption{\label{Fig:PD} ({\em Color online}). Phase diagram of the PNJL model.
The scale $T_c^0 = 173.9$ MeV corresponds to the critical temperature at
$\mu_5 = 0$. }
\end{center}\end{figure}

In Figure~\ref{Fig:PD} we plot the phase diagram of the
model in the $\mu_5 - T$ plane, for the case $\mu = 0$. At any value
of $T$ and $\mu_5$, the chiral condensate and the Polyakov loop
expectation value are computed by a minimization procedure of the full
potential~\eqref{eq:OB}. The structure of our phase diagram
is in agreement with previous model studies, see~\cite{Fukushima:2010fe,Chernodub:2011fr}.
Since chiral symmetry is broken explicitly
by the quark mass and the phase transitions are actually
crossovers, we identify the critical temperature with that at
which $dL/dT$ is maximum. We have checked that the latter deviates
from that at which $|d\sigma/dT|$ is maximum only of a few MeV, in
the whole range of parameters studied. With an
abuse of nomenclature, we dub the pseudo-critical lines as second
order and first order. It is clear
from the context that the
term second order transition has to be taken as a synonym of
smooth crossover; similarly, the term first order transition is a
synonym of discontinuous jump of the order parameters.

In the Figure~\ref{Fig:PD}, the grey dashed line
corresponds to a smooth crossover. The solid line, on the other hand, denotes the first
order transition. The dot corresponds to CP$_5$.  
In the PNJL model we have access
to the chiral condensate and to the Polyakov loop expectation
value. As a consequence, we can label the phases of the model in
terms both of confining properties, and of chiral symmetry. In the
model at hand, because of the entanglement in
Equation~\eqref{eq:Run}, the deconfinement and chiral symmetry
restoration crossovers take place simultaneously. The region below the pseudo-critical
line is characterized by confinement and spontaneous breaking of
chiral symmetry; we label this phase as the confinement phase. On
the other hand, the phase above the critical line is identified
with the Quark-Gluon-Plasma phase. In this case, CP$_5$ is both
{\it chiral} and {\it deconfinement} critical endpoint.
For what concerns the coordinates of CP$_5$ we find, for the PNJL model, 
\begin{equation}
\left(\frac{\mu_{5c}}{T_c^0},\frac{T_c}{T_c^0}\right) =
(1.73,0.96)~,~~~~~\text{CP}_5~\text{(PNJL)}~, \label{eq:CP5}
\end{equation}
where $T_c^0 = 173.9$ MeV is the deconfinement temperature at $\mu
= \mu_5 = 0$.

%\section{Critical endpoint at finite chemical potential}
Next we turn to discuss the more general case with both $\mu_5$
and $\mu$ different from zero. Our scope is to show that, at least
within the models, CP naturally evolves into CP$_5$. In particular the PNJL model, which is in
quantitative agreement with the Lattice at zero chemical
potential, gives a numerical relation among the coordinates of CP
and CP$_5$, which might be taken as a guide to estimate the
coordinates of CP in QCD, once CP$_5$ is detected.

\begin{figure}[t!]
\begin{center}
\includegraphics[width=7cm]{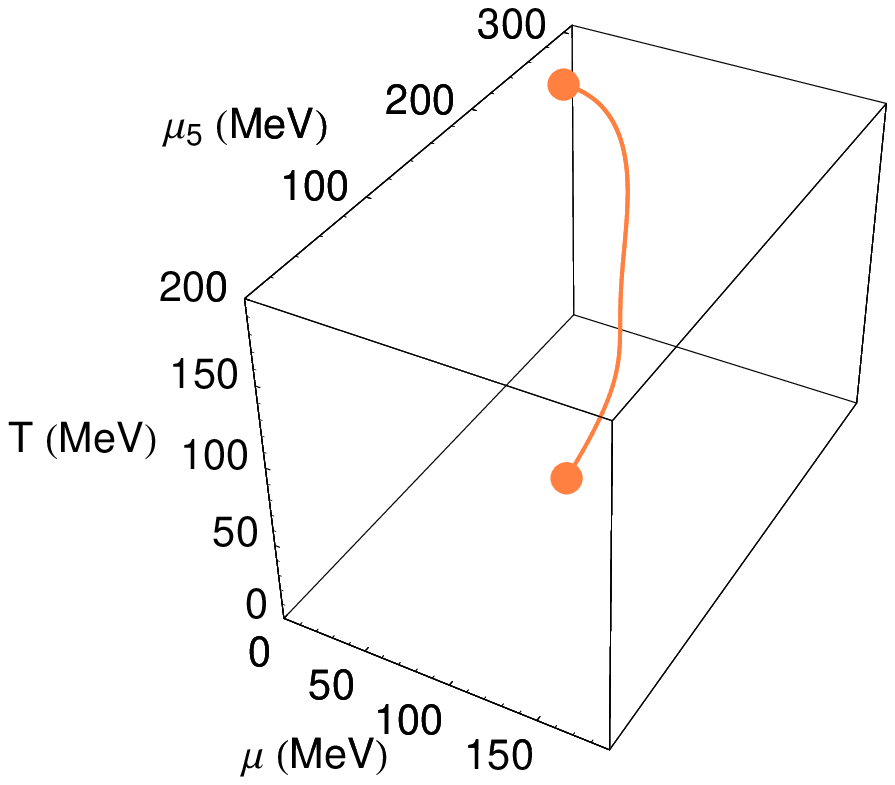}~~~\includegraphics[width=7cm]{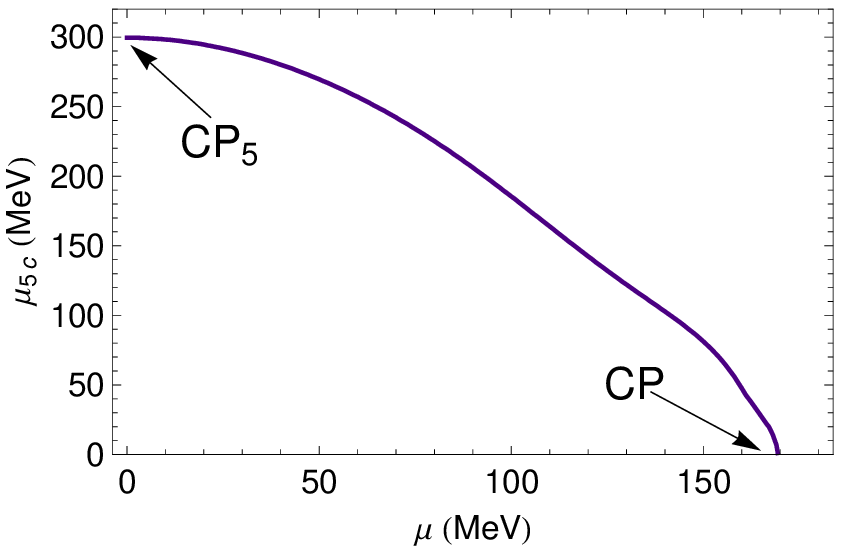}
%\end{center}
\caption{\label{Fig:EV} ({\em Color online}). Evolution of the
critical endpoint in the $\mu-\mu_5-T$ space, for the PNJL model.}
\end{center}\end{figure}

In Figure~\ref{Fig:EV} we collect our data on the critical point
of the phase diagram in the $\mu-\mu_5-T$ space, in the case of
the PNJL model. The
orange solid line is the union of the critical points computed
self-consistently at several values of $\mu$: at any value of
$\mu$, a point on the line corresponds to the critical point of
the phase diagram in the $\mu_5-T$ plane. Thus the line
pictorially describes the evolution of the critical point of the
chiral model at hand, from CP to CP$_5$. In the same Figure
we plot a projection of the critical endpoint evolution curve onto the $\mu - \mu_{5}$
plane, for the PNJL model. The indigo solid line corresponds to
the $\mu_5$-coordinate of the critical endpoint. The critical
temperature is not so much affected when we continue CP$_5$ to CP
(we measure a change approximately equal to the $3\%$), therefore
the projection in the $\mu-T$ plane is redundant.
Our numerical results suggest the following
relationships between the critical coordinates at $\mu=0$ and $\mu_5 = 0$:
\begin{equation}
\frac{\mu_c}{\mu_{5c}} \approx 0.53~,~~~\frac{T_c}{T_{5c}} \approx
0.97~,~~~~~~\text{(PNJL)}~. \label{eq:BBB}
\end{equation}

The model predictions~\eqref{eq:BBB} relate the coordinates of CP
to those of CP$_5$. In particular, it is interesting that the
critical temperature is almost unchanged in the continuation of CP
to CP$_5$. Of course, since these results are deduced by a model,
it is extremely interesting and important to study how
Equation~\eqref{eq:BBB} is affected by the value of the
bare quark mass, as well as by further interactions in 
the vector and axial-vector channels. 
These topics will
be the subject of a forthcoming publication~\cite{forth}.
It is worth to anticipate some of the results of~\cite{forth},
namely that a larger value of the quark mass, as well as the
vector interaction, move CP$_5$ to larger values of $\mu_5$.
The combination of these two factor, together with the finite size
of the lattice cell, might explain the absence of CP$_5$
in the Lattice simulations~\cite{Yamamoto:2011gk}.

\section{Conclusions and Outlook}
In this talk, we have reported on our results about the phase
structure of hot quark matter in presence of a background of chiral 
charge density. Such a background is introduced
by virtue of a chemical potential, $\mu_5$, conjugated to the
chirality imbalance, $n_5 = n_R - n_L$.
Because of the fluctuations of the topological charge,
which is connected with chirality imbalance in QCD via the 
quantum anomaly, $\mu_5$ should be treated as a pure mathematical
artifice, and cannot be considered as a true chemical potential.

This study is partly motivated by the potential
applications to the Chiral Magnetic Effect~\cite{Kharzeev:2007jp,Fukushima:2008xe,Buividovich:2009wi}. 
The latter might be relevant for the phenomenology of
heavy ion collisions, because a copious production of gluon configurations,
the QCD sphalerons, with a finite winding number is expected in the quark-gluon-plasma
phase of QCD, see~\cite{Moore:2010jd} and references therein. Because
of the chiral Ward identity, the interaction of the sphalerons 
with the quarks causes a chirality change of the latters. As a consequence,
a copious production of local domains in which chirality is imbalanced,
is expected in the quark-gluon-plasma.

After an overview on the phase diagram of hot quark matter
at finite $\mu_5$, obtained within an effective model,
we have suggested the possibility of continuation
of the critical endpoint of the phase diagram of $N_c = 3$ QCD,
CP, to a critical endpoint dubbed CP$_5$ at finite $\mu_5$ and $\mu=0$.
The worldsheet ${\cal W}_5\equiv\{\mu=0,\mu_5=0\}$ has
the merit that it can be simulated on the
Lattice~\cite{Fukushima:2008xe,Yamamoto:2011gk} for $N_c = 3$. 
Even if Lattice results have been already published~\cite{Yamamoto:2011gk}, 
more care should be taken in the derivation of the relation 
between $n_5$ and $\mu_5$, since $n_5$ is a nonconserved quantity,
hence it suffers renormalization effects which should be taken
into account.

The phase structure that we have discussed here is based on 
the PNJL model with entanglement vertex,
introduced in~\cite{Sakai:2010rp}, which offers a description of
the QCD thermodynamics in terms of collective degrees of freedom,
which is in quantitative agreement with Lattice data at zero and
imaginary chemical potential.

One of our ideas is that simulations in the worldsheet ${\cal W}_5$ 
might reveal the
existence of a critical endpoint, CP$_5$, in the phase diagram.
Then, this critical point might be interpreted as the continuation
of the critical point which is expected to belong to the phase
diagram of real QCD, because of the continuity summarized in
Fig.~\ref{Fig:EV}.  Hence it would be an indirect evidence of the
existence of the critical point in real QCD. 

In our calculations there are some factors that we have not
included for simplicity, and that affect the location of CP$_5$. 
For example, the bare quark mass and the vector
interaction move CP$_5$ to higher values of $\mu_5$. 
These observations might be helpful to understand why in the Lattice 
simulations of~\cite{Yamamoto:2011gk}, no critical endpoint is detected.
We plan to report on the aforementioned topics in the next future~\cite{forth}.

\vspace{1cm}
{\bf Acknowledgements}. It is a pleasure to thank the organizers
of the {\em Eleventh Workshop on Non-Perturbative
Quantum Chromodynamics} for their kind invitation. Part of this work was inspired by
stimulating discussions with H.~Warringa, who is acknowledged.
Moreover, we acknowledge M.~Chernodub, M.~D'Elia, P.~de Forcrand, G. Endrodi, M.~Frasca,
K.~Fukushima,
R.~Gatto, T. Z. Nakano, A.~Ohnishi, O. Philipsen, Y. Sakai, T. Sasaki, A.~Yamamoto and N.~Yamamoto for
correspondence and many discussions on the topics presented here. 
This work was supported by the Japan Society for the
Promotion of Science under contract number P09028.

%\section*{References}

\end{document}